# On long-term variations of solar wind parameters and solar activity

V.I. Vlasov,  R.D. Dagkesamanskii, V.A. Potapov, S.A. Tyul'bashev, I.V. Chashei

Lebedev Physical Institute RAS, Moscow

**Abstract**

Comparison is carried out of the long term variation of the year averaged solar wind speed and interplanetary scintillation index with the variations of Wolf's numbers and $A_P$ indexes of geomagnetic activity for the data of 20-24 solar activity cycles. It is shown that the slow non-monotonous trend in the scintillation parameters at middle and high heliolatitudes exists with the typical scale of order of century cycle. Correlation between the variations of Wolf's numbers and anomalies of the air temperature is analyzed for long data series from 1610 up to the present time. Possible application of the results to the global climate problem is discussed.

**Introduction**

Interplanetary scintillations are caused by the diffraction of radio waves on moving inhomogeneities of the solar wind [1]. Observations of interplanetary scintillations make it possible to determine the parameters of the solar wind in a wide range of heliocentric distances and solar latitudes, as well as to study their evolution in the solar activity cycle [1] and over longer time intervals [2-4]. In this paper, based on the accumulated observational data, variations in the parameters of interplanetary scintillations over a long time interval covering the last 4 cycles of solar activity are considered. These data allow us to identify a non-monotonous trend in changes in the parameters of the solar wind, a comparison of this trend with the cycle of solar activity is carried out. Comparison of similar trends in archival data on solar activity with changes in the global climate of the Earth allows us to make a forecast about possible climatic trends, assuming the determining role of the influence of the Sun.

**Long-term variations of interplanetary scintillation parameters**

Next, variations of two main parameters of interplanetary scintillation will be considered: the index of scintillation and the velocity of the diffraction pattern. The scintillation index $m$ is defined as the relative fluctuations of the radiation flux of a transmission radio source with a characteristic time scale of the order of 1 second [1]. It is usually assumed that this parameter is proportional to the average concentration of solar wind plasma along the line of sight [1]. The speed of movement of the scintillation pattern $V$ is measured by simultaneous observations on radio telescopes spaced over a distance of about hundreds of kilometers. This parameter approximately coincides with the solar wind speed averaged over the visual beam [4]. The second and third panels of Fig. 1 show the measurement data averaged over annual intervals of the $m$ and velocity $V$ for a period of about 40 years (20-24 cycles of activity) for the same radio source 3C 48, which shines through the solar wind in the region of solar latitude 20º - 70º. The scintillation indices were measured by the Pushchino Radio Astronomy Observatory (PRAO) of the Lebedev Physical Institute (LPI), all values are given at a frequency of 100 MHz. Data on the solar wind speed were obtained at a frequency of 327 MHz on the three-point system of Nagoya University, Japan (https://www.isee.nagoya-u.ac.jp/en/ ). The first panel of Fig. 1 shows the Wolf numbers $W$, characterizing the level of solar activity. A comparison of the data from the second and third panels of Fig. 1 shows that within each activity cycle there is an anticorrelation between the $m$ and $V$: at the maximum of activity, the speed is minimal, and the scintillation index is maximal; at the minimum of activity, the speed is maximal, and the scintillation index is minimal. This dynamics is due to the fact that the minimum activity at high latitudes is dominated by a fast rarefied solar wind, and at low latitudes by a denser slow wind; the maximum activity at all latitudes is dominated by a slow solar wind [1,4]. Against the background of cyclical variations, there is a non-monotonic slow trend, of the same nature for the speed and the scintillation index. The maximum of the trend in the parameters of scintillation is about the duration of one cycle late in relation to the maximum in the level of solar activity. On the fourth panel, Fig. 1 The $A_p$ values of the geomagnetic activity index are given (http://www.swpc.noaa.gov/products/solar-cycle-progression ), the slow trend of which, as can be seen, correlates with a similar trend in the parameters of interplanetary scintillations.

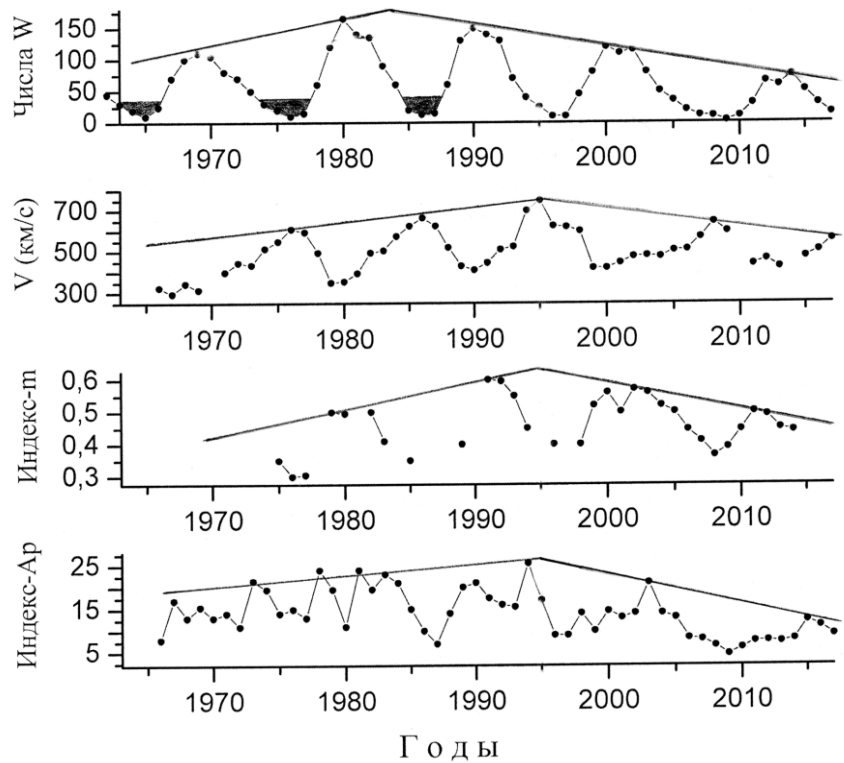

Fig.1 Series of average annual values of the Wolf numbers $W$, values of the solar wind velocity $V$ and the scintillation index $m$ (according to radio astronomy data) and the geomagnetic activity index $A_p$.

In [5], data on the magnetic field, concentration and speed of the solar wind are presented according to measurements on the *Wind* satellite, also in 20-24 cycles of activity. In these data, both cyclical dynamics and a slow trend are absent for the specified parameters of the solar wind. The *Wind* data [5] do not contradict the data of radio astronomy observations, since the satellite was located in the region of low solar latitudes, near the plane of the solar equator, while the data of the second and third panels of Fig. 1 refer to medium and high solar latitudes. Therefore, it can be argued that both the cyclical dynamics and the slow trend in the solar wind are due to variability at medium and high solar latitudes. In terms of cyclical dynamics, this conclusion is also consistent with measurements on the *Ulysses* spacecraft during its flights through the equatorial regions and over the poles at the minimum and maximum of solar activity.

**About a possible application to the problem of global climate**

In recent years, the problem of global temperature rise of the Earth's atmosphere has been intensively discussed in the scientific literature and in the media. Based on archival data, we will consider a possible connection between global temperature and long-term cyclical changes in solar activity. The first instrumental measurements of Wolf numbers at the Zurich Observatory date back to 1610. The top panel of Fig.2 shows the longest available series of Wolf numbers. The nonmonotonic trend in Fig.1 is represented by solid straight lines drawn as an envelope of the maxima of 11-year cycles from the 20th to the 24th. The level predicted by extrapolation of the declining linear trend of the beginning of the 25th cycle is shown in the dotted line. The fragment of the graph from the work [7] presented in the lower panel of Fig. 2 corresponds to the values of anomalies of the surface air temperature before 1980, reconstructed according to the data of *18O*. The dots on the bottom panel of Fig. 2 show the values of the temperature anomaly in the current period, the data are borrowed from the website (*http://www.climate4you.com*). The two previous secular minima in terms of Wolf numbers and air temperature are connected by dashed lines, which, as can be seen from Fig. 2, turn out to be parallel, and there is a delay in the global temperature with respect to the Wolf numbers. It can be assumed that the relationship between temperature and Wolf numbers persists in the current cycle, then the maximum temperature will correspond with some delay to the maximum in Wolf numbers. The minimum activity in the current secular cycle has not yet been reached, but the decreasing envelope in the right part of the upper panel of Fig.2 suggests that already during the beginning of the 25th cycle, that is, within 10 years, we can expect a change in the phase of global temperature growth to the phase of decline.

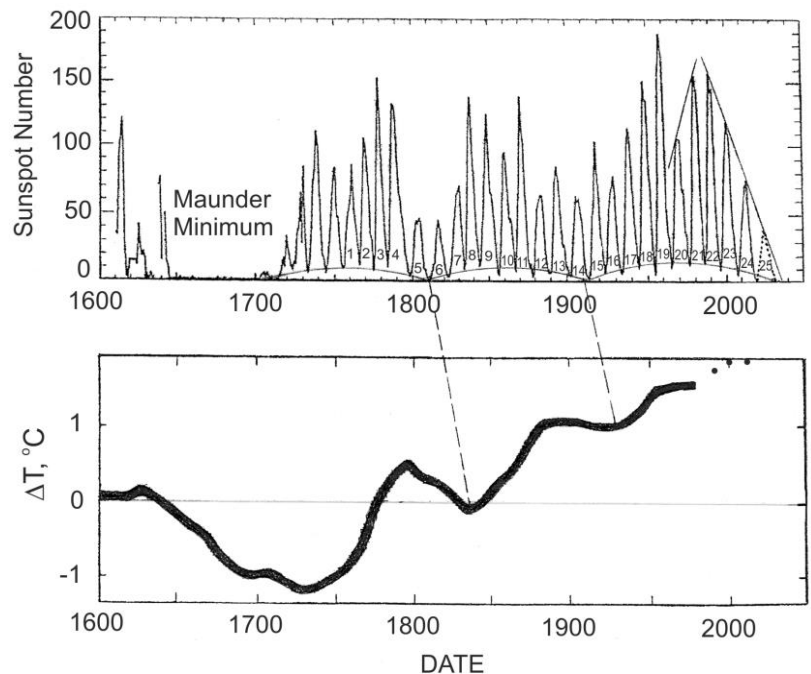

Fig.2 Comparison of the temperature anomaly $\Delta T$ reconstructed according to *18O* data in continental Siberia (Altai Krai, Belukha Glacier) with the number of sunspots (Wolf numbers) in the time interval 1610-2019.

We assume, like many other authors, that variations in global temperature are determined by age-old changes in the level of solar activity, that is, by natural causes. To test this hypothesis, we conducted a comparative analysis of the variation in the average concentration of atmospheric $CO_2$ (*www.climate4you.com/GreenhouseGasses.htm*) and global temperature (*www.climate4you.com/GlobalTemperatures.htm*). The cross-correlation of the two time series shows that $CO_2$ variations are lagging relative to temperature variations with a lag time of about one year. A similar estimate of the delay of carbon dioxide relative to temperature was also obtained in [8].

## Conclusion

The results of observations of interplanetary scintillations covering a period of about 4 cycles of solar activity show that there is a slow non-monotonic trend in the solar wind against the background of cyclic variations. Since there is no similar variability in the near-Earth plasma of the solar wind, it can be concluded that both the cyclic variability and the slow trend are due to changes in the mid-latitude and high-latitude regions of the solar corona. Unlike cyclic variability, for which the plasma velocity and concentration vary counterphase, for slow variability, the velocity and concentration change in a similar way. Cyclic variability is known to be associated with the appearance of polar coronal holes in the minimum of activity. It can be assumed that the slow variability is somehow connected with the secular modulation of the magnetic structure of the solar atmosphere. The slow trend of the solar wind is delayed by about the duration of the solar cycle in relation to solar activity. Analysis of the possible connection of slow variations in the global temperature of the Earth's atmosphere with solar activity allowed us to make an assumption, in our opinion, reasonable, about the transition from warming to cooling within the next 10 years. Therefore, in the foreseeable future it will be possible to make a choice between the natural and anthropogenic nature of global warming.